\documentclass[aps,prb,amsfonts,amssymb,twocolumn,amsmath,preprintnumbers,nofootinbib,floatfix,
showpacs]{revtex4-1}
\usepackage[dvips]{graphics}
\usepackage{graphicx}
\usepackage{bm}
\begin{document}

\title{Time-reversal symmetry breaking state near the surface of $s_{\pm}$-superconductor}
\author{A. M. Bobkov}
\affiliation{Institute of Solid State Physics, Chernogolovka,
Moscow reg., 142432 Russia}
\author{I. V. Bobkova}
\affiliation{Institute of Solid State Physics, Chernogolovka,
Moscow reg., 142432 Russia}

\date{\today}

\begin{abstract}

The structure of superconducting order parameter near the surface of a two-band superconductor
with $s_{\pm}$ order parameter in the bulk is theoretically investigated.
The main parameter of the surface, which determines
the appropriate physics is the coefficient of the interband scattering $R_{12}$.
For small $R_{12}$ the superconducting order parameter is only suppressed
to some extent near the surface for the both bands.
For intermediate and strong interband scattering there are two possible non-trivial surface
states of the order parameter: (i)
purely real solution, where the symmetry of the superconducting state near the
surface is changed from $s_{\pm}$ to conventional $s_{++}$ and (ii) time-reversal symmetry breaking (TRB) state.
In this state the order parameters in the two bands acquire
phases $\phi_{1,2}(x) \neq (0,\pi)$ upon approaching the surface.
We argue that at low temperatures the TRB surface state can be more energetically favorable
than the $s_{\pm} \to s_{++}$ time reversal symmetry conserving state (TR).
For higher temperatures up to $T_c$ only the TR state can exist. The transition between
the two temperature regions is rather sharp. Signatures of the transition between the TRB
and the TR surface states can be detected by the measurements of the local density of states
and the angle-resolved density of states.

\end{abstract}

%%% PACS numbers

\maketitle

\section{introduction}

The discovery of a new family of iron-based high-temperature superconductors with
distinct multi-orbital band structure \cite{bruning08,zhao08,pourovskii08}
has renewed interest to the problem of multi-band superconductivity, firstly
discussed fifty years ago \cite{suhl59,moskalenko59}. In iron-pnictide \cite{inosov10} and
iron chalcogenides \cite{argyriou10,li10,lumsden10,qiu09} the Fermi surface (FS) has two hole-pockets
around $\Gamma=(0,0)$  and two electron-pockets around $M=(\pi,\pi)$. The proximity of
superconductivity to an antiferromagnetic phase
suggests that the magnetic fluctuations play an important role
in the understanding of pairing mechanism. It was proposed
\cite{mazin08,kuroki08} that in this case the so-called $s_{\pm}$ superconducting
state can be realized. It is characterized by the phase difference $\pi$ between the
superconducting condensates arising on the hole Fermi surfaces around
$\Gamma$ point and the electron Fermi surfaces
around M point. This state has been favored by a variety of models within random phase approximation
(RPA) \cite{kuroki08,cvetkovic09_1,graser09} and renormalization group techniques
\cite{chubukov08,wang09,cvetkovic09_2}.

The problem of inhomogeneities, such as disorder, surfaces and interfaces, is substantially
more complicated in these materials than in one-band systems and contains rich new physics,
which is explored insufficiently by now. For simplicity we consider further only two Fermi surfaces
(one electron FS and one hole FS). An incoming quasiparticle
belonging to one Fermi surface (we also use term "band" below) can be scattered by the impurity
(surface, interface) as into the same band (intraband scattering), so as into
the other band (interband scattering). Interplay of the interband scattering with the different signs
of the superconducting order parameter (OP) on different Fermi surfaces can lead
to various new physical phenomena. For example, it has been
shown that in the $s_{\pm}$ state any fully {\it interband} nonmagnetic impurity suppresses $T_c$
in the same way as a magnetic impurity in a single band superconductor \cite{golubov97}. However, a much
slower pairbreaking rate can be achieved if one assumes that the scattering is primarily intraband rather
than interband. In the pure intraband scattering limit Anderson's theorem can be applied and, consequently,
no $T_c$ suppression occurs. The rate of $T_c$ suppression is, therefore, very sensitive to the relative
strength of the intra- and interband scattering rates. Further, it was shown \cite{efremov11} that
depending on the ratio of inter- to intraband pairing constants there are two possible types of $s_{\pm}$
superconductivity. For the first one $T_c$ is suppressed as disorder is increased and vanishes at
a critical value of the scattering rate. For the second type $T_c$ tends to a finite value upon
increasing of disorder, while the order parameter undergoes a transition from $s_{\pm}$ to a gapless state
and then to a fully gapped $s_{++}$ state.

Surface and interface phenomena in $s_{\pm}$-superconductors have also attracted
considerable recent attention. The formation of bound states at a free surface of an $s_{\pm}$-superconductor
\cite{ghaemi09,nagai09_1,onari09,nagai09_2},
at an S$_\pm$/N \cite{choi08,linder09,golubov09,sperstad09}, an N/S/S$_\pm$ junction \cite{feng09}
and at Josephson junctions including $s_{\pm}$-superconductors \cite{tsai09,sperstad09} was investigated theoretically.
In particular, the finite energy subgap bound states were found
and their influence on the conductance spectra and Josephson current was investigated. It was found that
the particular energies of the bound states are very sensitive to the relative strengths of the intraband and interband
scatterings.

Further, it was found \cite{bobkov10} that the spatial behavior of the OP
at a surface of a $s_{\pm}$ superconductor can not be reduced to a trivial suppression.
If the interband scattering at a surface
is of the order of the intraband one or dominates it, the symmetry of the superconducting state near the
surface can be changed from $s_{\pm}$ to conventional $s_{++}$.
The spatial region of existing of this surface
conventional superconductivity is very sensitive to the relative values of interband and intraband
pairing potentials. The transition between the bulk $s_{\pm}$ an surface $s_{++}$ superconductivity
regions occurs through a region of suppressed order parameter, to some extent analogously to the discussed
above gapless superconductivity in the disorder problem.

In the present paper we continue investigation of the OP spatial behavior
near a surface of a $s_{\pm}$ superconductor. It is found that besides the discribed above
purely real solution for the order parameter there is another solution, which starts from the same
$s_{\pm}$ state in the bulk, but manifests qualitatevely different behavior near the surface.
This state is time-reversal symmetry breaking (TRB): upon approaching the surface the order parameters
in the two bands (called 1 and 2) acquire phases $\phi_{1,2}(x) \neq (0,\pi)$, while the bulk
values $\phi_1 = 0$ and $\phi_2 = \pi$ restore within a few superconducting coherence
lengths from the surface. We argue that this TRB surface state can be more energetically favorable
than the discussed above $s_{\pm} \to s_{++}$ time reversal symmetry conserving state (TR). The TRB
surface state only occurs at low temperatures, while for higher temperatures up to $T_c$ only
the TR state can exist. The transition between the two temperature regions is rather sharp.
Further, we show that the local density of states (LDOS) spectra are very sensitive to the particular surface
state of the order parameter and, therefore, experience a sharp change when temperature crosses
the transition point. Therefore, the transition between TRB and TR surface states can be detected
by investigation the LDOS.

The possibility of TRB state has been already discussed in literature \cite{lee09,agterberg99,ng09,stanev10,
tanaka10,hu11,platt11,ota11,lin11,garaud11,carlstrom11,stanev11}. However, the physical mechanism,
underlying this TRB state is a frustration between three or more Josephson-coupled bands.
On the contrary, we consider only two Fermi pockets and the surface TRB state is generated by the interband
scattering between them.

The paper is organized as follows. The model system we study and our theoretical method are described
in Sec.~\ref{model}. In Sec.~\ref{OP} we represent the results of the self-consistent calculations of
the superconducting order parameter for our surface problem and discuss the TRB state. In Sec.~\ref{bs}
the LDOS is calculated and its dependence on temperature is analyzed. Our conclusions are given
in Sec.~\ref{conc}.

\section{model and method}

\label{model}

We consider an impenetrable surface of a clean two-band superconductor. The OP
is assumed to be of $s_\pm$-symmetry in the bulk of the superconductor, that is the phase difference
between the OP's in the two bands is $\pi$. It is supposed that an incoming quasiparticle
from band 1,2 can be scattered by the surface as into the same band (intraband scattering), so as into
the other band (interband scattering). In order to maximize the effects generated by the surface
scattering we assume that the $c$-axis is parallel to the surface. 

We make use of the quasiclassical theory of superconductivity, where all the relevant physical
information is contained in the quasiclassical Green function
$\hat g_i(\varepsilon, \bm p_f, x)$ for a given quasiparticle trajectory.
Here $\varepsilon$ is the quasiparticle energy measured
from the chemical potential, $\bm p_f$ is the
momentum on the Fermi surface (that can have several
branches), corresponding to the considered trajectory,
$x$ is the spatial coordinate along the normal to the surface and
$i=1,2$ is the band index. Quasiclassical Green function
is a $2 \times 2$ matrix in  particle-hole space, that is denoted by the symbol $\hat ~$. The equation
of motion for $\hat g_i(\varepsilon, \bm p_f, x)$ is the Eilenberger equation
subject to the normalization condition \cite{larkin69,eilenberger68}. For superconductivity of $s_\pm$-type
electrons from different bands cannot form a pair. Therefore, the Eilenberger equations
corresponding to the bands 1 and 2 are independent. The trajectories belonging to the different bands
can only be entangled by the surface, which enters the quasiclassical theory in the form of effective
boundary conditions connecting the incident and outgoing trajectories.

However, owing to the normalization
condition for the quasiclassical propagator, the boundary
conditions for the quasiclassical Green functions are formulated
as non-linear equations \cite{shelankov84,zaitsev84,millis88}.
For the multiband case, considered here, they are rather complicated and their practical use is limited.
For this reason in the present work we make use of the quasiclassical formalism
in terms of so-called Riccati amplitudes \cite{shelankov80,eschrig00}, which allows an explicit formulation
of boundary conditions \cite{eschrig00,shelankov00,fogelstrom00,zhao04,eschrig09}. The retarded Green function
$\hat g_i(\varepsilon, \bm p_f, x)$,
which is enough for a complete description of an equilibrium system, can be parametrized via two
Riccati amplitudes (coherence functions) $\gamma_i(\varepsilon, \bm p_f, x)$ and
$\tilde \gamma_i(\varepsilon, \bm p_f, x)$ (in the present paper we follow
the notations of Refs.~\onlinecite{eschrig00,eschrig09}).
The coherence functions obey the Riccati-type transport equations.
In the considered here case of two-band clean $s_\pm$-superconductor the equations for the two bands are independent
and read as follows
\begin{equation}
iv_{ix}\partial_x \gamma_i+2\varepsilon \gamma_i =-\Delta_i^*(x) \gamma_i^2-\Delta_i(x)
\label{gamma}
\enspace ,
\end{equation}
\begin{equation}
\tilde \gamma_i (\varepsilon, \bm p_f, x)=\gamma_i^*(-\varepsilon, -\bm p_f, x)
\label{tilde_gamma}
\enspace .
\end{equation}
Here $v_{ix}$ is the normal to the surface Fermi velocity component for the quasiparticle belonging to band $i$.
$\Delta_i(x)$ stands for the OP in the $i$-th band and should be found self-consistently.

Let us suppose that the surface is located at $x=0$ and the superconductor occupies the halfspace $x>0$.
For the sake of simplicity we assume that the surface is atomically clean and, consequently,
conserves parallel momentum component. Then there are four quasiparticle trajectories, which are involved
in each surface scattering event. These are two incoming trajectories belonging to the bands 1,2 (with $v_{ix}<0$)
and two outgoing ones (with $v_{ix}>0$). It can be shown \cite{eschrig00,eschrig09} that the coherence function
$\gamma_i(\varepsilon, \bm p_f, x)$, corresponding to the incoming trajectory
can be unambiguously calculated making use of Eq.~(\ref{gamma})
and starting from its asymptotic value in the bulk
\begin{equation}
\gamma_i^{b}=-\frac{\Delta_i^b{\rm sgn}\varepsilon}{|\varepsilon|+\sqrt{(\varepsilon+i\delta)^2-{\Delta_i^b}^2}}
\label{gamma_bulk}
\enspace ,
\end{equation}
where $\Delta_i^b$ is the bulk value of the OP in the appropriate band, $\delta>0$ is an infinitesimal.
The coherence function
$\tilde \gamma_i(\varepsilon, \bm p_f, x)$ is determined unambiguously by the asymptotic conditions for
the outgoing trajectories and can be obtained according to Eqs.~(\ref{gamma}),(\ref{tilde_gamma}).

Otherwise, the coherence functions $\gamma_i(\varepsilon, \bm p_f, x)$ for the outgoing trajectories
and, correspondingly, $\tilde \gamma_i(\varepsilon, \bm p_f, x)$ for the incoming ones should be calculated
from Eq.~(\ref{gamma}) supplemented by the boundary conditions at the surface and Eq.~(\ref{tilde_gamma}).
The surface is described by
the normal state scattering matrix for particle-like excitations, denoted by $S$ and
for hole-like excitations, denoted by $\widetilde S$.
The scattering matrix $S$ have elements $S_{\bm k_i\bm p_j}$,
which connect outgoing quasiparticles from band $i$ with momentum $\bm k_i$ to the incoming ones
belonging to band $j$ with momentum $\bm p_j$.
Here and below all the momenta corresponding to the incoming trajectories are denoted by letter $\bm p$
and all the momenta for the outgoing quasiparticles are denoted by $\bm k$. For the model we consider $S$
is a $2 \times 2$-matrix in the trajectory space
(for the particular value of the momentum component parallel to the surface, which is denoted by $\bm p_{||}$). 
It obeys the unitary condition $SS^\dagger=1$. We also assume that the surface hamiltonian by itself posesses
the time-reversal symmetry, which imposes an additional constraint on the scattering matrix 
$S(\bm p_{||})=S^{tr}(-\bm p_{||})$ \cite{millis88}. In the absence of spin-orbit interaction the $S$-matrix
elements are only functions of $|\bm p_{||}|$. Then, without loss of generality the scattering matrix
can be parameterized by three quantities $R_{12}$, $\Theta$ and $\alpha$ as follows
\begin{equation}
\left(
\begin{array}{cc}
S_{\bm k_1\bm p_1} & S_{\bm k_1\bm p_2} \\
S_{\bm k_2\bm p_1} & S_{\bm k_2\bm p_2}
\end{array}
\right)=
\left(
\begin{array}{cc}
\sqrt R_0 e^{i\Theta} & i \alpha \sqrt{R_{12}} \\
i \alpha \sqrt{R_{12}} & \sqrt R_0 e^{-i\Theta}
\end{array}
\right)
\label{S_electron}
\enspace ,
\end{equation}
where $R_0$ and $R_{12}$ are coefficients
of intraband and interband reflection, respectively. They obey the constraint $R_0+R_{12}=1$.
The sign factor $\alpha=\pm1$ and the phase factor $\Theta$ do not enter the boundary conditions and, 
consequently, do not influence all the obtained results.
While in general the scattering matrix elements are functions of $|\bm p_{||}|$, 
we disregard this dependence in order to simplify the analysis.
The scattering matrix $\widetilde S$ for hole-like excitations is connected to $S$ by the relation
$\widetilde S(\bm p_{||})=S^{tr}(-\bm p_{||})$, that is in the case we consider $\widetilde S=S$.

From the general boundary conditions \cite{eschrig09}, which are also valid
for a multiband system, one can obtain the explicit values of the coherence functions
$\gamma_i (\varepsilon, \bm k, x=0)$ and $\tilde \gamma_i(\varepsilon, \bm p, x=0)$
via the scattering matrix elements and the values of the coherence functions
$\gamma_i (\varepsilon, \bm p, x=0)$ and $\tilde \gamma_i(\varepsilon, \bm k, x=0)$
at the surface. They read as follows
\begin{equation}
\gamma_{1\bm k}=R_0 \gamma_{1\bm p}+R_{12} \gamma_{2\bm p}-\frac{R_0R_{12}\tilde \gamma_{2\bm k}
(\gamma_{1\bm p}-\gamma_{2\bm p})^2}
{1+\tilde \gamma_{2\bm k}\left( R_{12} \gamma_{1\bm p}+R_0 \gamma_{2\bm p} \right)}
\label{gamma_surface}
\enspace ,
\end{equation}
\begin{equation}
\tilde \gamma_{1\bm p}=R_0 \tilde \gamma_{1\bm k}+R_{12} \tilde \gamma_{2\bm k}-\frac{R_0R_{12}
\gamma_{2\bm p}(\tilde \gamma_{1\bm k}-\tilde \gamma_{2\bm k})^2}
{1+\gamma_{2\bm p}\left( R_{12} \tilde \gamma_{1\bm k}+R_0 \tilde \gamma_{2\bm k} \right)}
\label{tilde_gamma_surface}
\enspace .
\end{equation}
Here the arguments $(\varepsilon,x=0)$ of all the coherence functions are omitted for brevity,
$\gamma_{i\bm p} \equiv \gamma_i(\bm p)$ and $\tilde \gamma_{i\bm p} \equiv \tilde \gamma_i(\bm p)$
and the analogous notations are used for $\gamma_i(\bm k)$ and $\tilde \gamma_i(\bm k)$.
Quantities $\gamma_{i\bm p}$ and $\tilde \gamma_{i\bm k}$, entering Eqs.~(\ref{gamma_surface})
and (\ref{tilde_gamma_surface}) are to be calculated from Eqs.~(\ref{gamma}) and (\ref{tilde_gamma})
supplemented by the appropriate asymptotic condition.
The coherence functions $\gamma_{2\bm k}$ and $\tilde \gamma_{2\bm p}$ are obtained by the interchanging
$1 \leftrightarrow 2$ in all the coherence function band indices at the right-hand side of
Eqs.~(\ref{gamma_surface}) and (\ref{tilde_gamma_surface}), respectively.

Now, substituting the coherence functions into the self-consistency equation
\begin{equation}
\Delta_i(x)=-T\sum \limits_{\varepsilon_n,j}\lambda_{ij}\left \langle
\frac{-2 i \pi \gamma_{j\bm p_f}}{1+\gamma_{j\bm p_f}\tilde \gamma_{j \bm p_f}} \right \rangle_{\bm p_f}
\label{self_consistency}
\enspace ,
\end{equation}
we iterate system (\ref{gamma})-(\ref{gamma_bulk}), (\ref{gamma_surface})-(\ref{self_consistency})
until it converges. In Eq.~(\ref{self_consistency}) $\lambda_{ii}<0$ is the dimensionless
pairing potential for band $i$ and $\lambda_{12}$, $\lambda_{21}$ are the dimensionless interband
pair-scattering potential. For simplicity we assume that the normal state densities of state coincide
for the two FS: $N_{F,1}=N_{F,2}$.
In this case $\lambda_{12} = \lambda_{21}$.  We choose $\lambda_{12}>0$, which stabilizes $s_\pm$ OP in the bulk.
The Matsubara frequencies $\varepsilon_n$ enter the coherence functions via the substitution
$\varepsilon+i\delta \to i \varepsilon_n$. $\langle ... \rangle_{\bm p_f}$ means the anomalous Green function
averaged over the entire Fermi surface, that is $\bm p_f$ incorporates as the incoming trajectories
$\bm p$, so as the outgoing ones $\bm k$. For concreteness  we suppose the Fermi surface to
be cylindrical for the each band.
However, our results do not qualitatively 
sensitive to the described above simplifying assumptions.

\section{Self-consistent order parameter: TRB state}

\label{OP}

\begin{figure}[!tbh]
  %\centerline{\includegraphics[clip=true,width=3.1in]{fig3.eps}}
            %\centerline{\includegraphics[clip=true,width=2.5in]{fig1b.eps}}
   \begin{minipage}[b]{\linewidth}
     \centerline{\includegraphics[clip=true,width=2.7in]{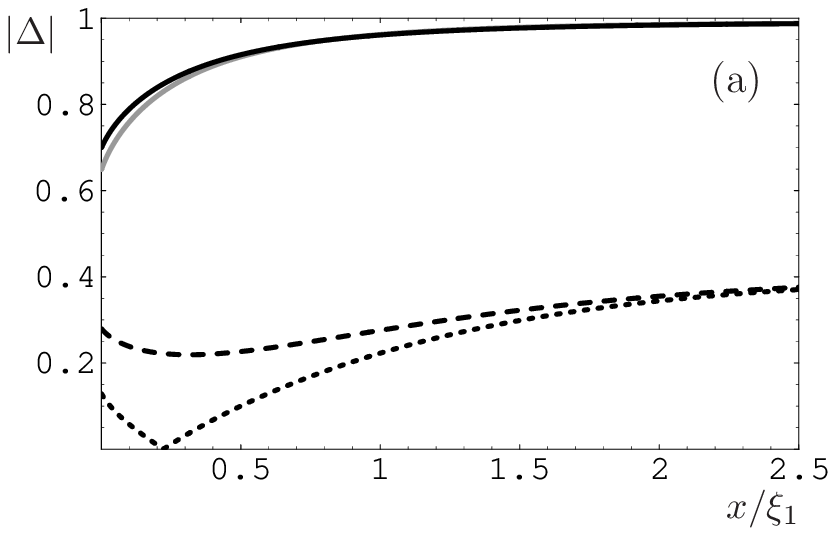}}
     \end{minipage}\hfill
    \begin{minipage}[b]{\linewidth}
   \centerline{\includegraphics[clip=true,width=2.7in]{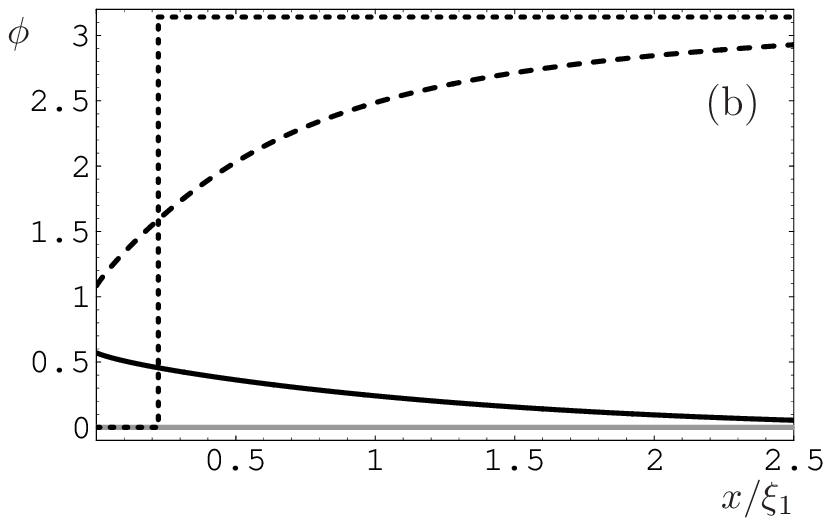}}
  \end{minipage}
   \caption{The OP profiles as functions of the spatial coordinate $x$ at $T=0.34$.
Panel (a) shows the OP absolute values: $|\Delta_1^{TRB}|$ (black solid line);
$|\Delta_2^{TRB}|$ (dashed); $|\Delta_1^{TR}|$ (gray solid);
$|\Delta_2^{TR}|$ (dotted).   
The OP phases are demonstrated in panel (b): $\phi_1^{TRB}$ (black solid);
$\phi_2^{TRB}$ (dashed); $\phi_1^{TR}$ (gray solid);
$\phi_2^{TR}$ (dotted).   
 $R_{12}=0.5$. In this Fig.
and below throughout the paper $\lambda_{11}=0.2771$,
$\lambda_{22}=0.2241$, and $\lambda_{12}=-0.004$; all the spatial coordinates are measured in units
of $\xi_1$; all the energies, including the OP absolute values, are measured in units of $\Delta_1^b(T=0)$
and temperature is measured in units of $T_c$.}
\label{Delta_compare}
\end{figure}

The spatial profiles of the OP calculated according to the described above technique at $T=0.34T_c$,
are represented in Fig.~\ref{Delta_compare}. We assume that
in the bulk $|\Delta_1^b|>|\Delta_2^b|$ and look for the solutions in the form $\Delta_{1,2}(x)=
|\Delta_{1,2}(x)|\exp [i\phi_{1,2}(x)]$. Panel (a) of Fig.~\ref{Delta_compare}
demonstrates the spatial profiles of OP absolute values for bands $1$ and $2$, while the corresponding
phases are plotted in panel (b). It is seen from the Figure that at this temperature the self-consistency
equation has two different solutions. One of them is the purely real (TR) solution, where the transition
from $s_{\pm}$ to $s_{++}$ state at a distance $\sim 0.25 \xi_1$ takes place.
Here $\xi_1=v_{1}/\Delta_1^b(T=0)$ is the superconducting coherence length for band 1.
This solution is characterized by going to zero $|\Delta_2(x)|$
and by abrupt jump of the phase $\phi_2(x)$ from $\pi$ to $0$. The other solution is time-reversal symmetry
breaking. $|\Delta_2(x)|$ does not approach zero, while $\Delta_{1,2}$ acquire
phases $\phi_{1,2}(x) \neq (0,\pi)$ in some region near the surface. The conjugated solution characterized
by $-\phi_{1,2}(x) $ is always exists, but is not represented in Fig.~\ref{Delta_compare}.

Due to the presence of the spatially dependent OP phase the TRB state is accompanied by the currents,
flowing perpendicular to the surface and transferred by electrons from the both bands. However, the currents,
transferred by the electrons from band 1 and from band 2 exactly compensate each other at any distance
$x$ from the surface. As a result, the total current, generated by the spatially dependent OP phase
in the TRB state, is zero, as it should be due to the charge conservation.

\begin{figure}[!tbh]
  %\centerline{\includegraphics[clip=true,width=3.1in]{fig3.eps}}
            %\centerline{\includegraphics[clip=true,width=2.5in]{fig1b.eps}}
   \begin{minipage}[b]{\linewidth}
     \centerline{\includegraphics[clip=true,width=2.7in]{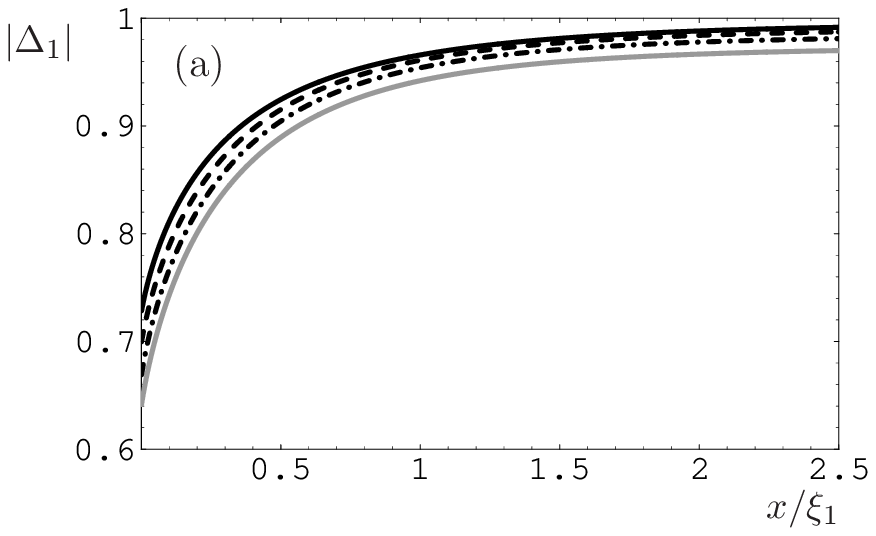}}
     \end{minipage}\hfill
    \begin{minipage}[b]{\linewidth}
   \centerline{\includegraphics[clip=true,width=2.7in]{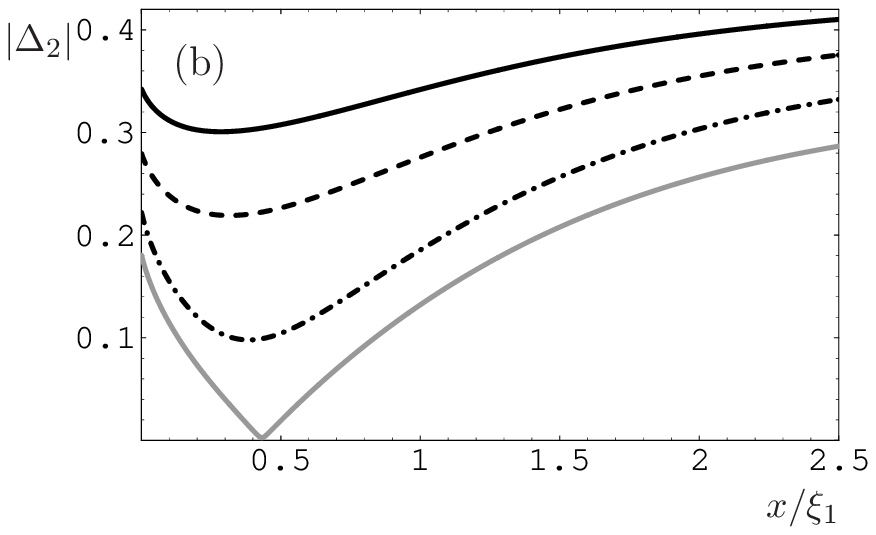}}
  \end{minipage}
   \caption{The OP absolute values for the TRB state as functions of the spatial coordinate $x$.
Panel (a) shows $|\Delta_1|$ and panel (b) represents $|\Delta_2|$.
For the both panels
different curves correspond to different temperatures: $T=0.30$ (black solid), $T=0.34$ (dashed),
$T=0.38$ (dashed-dotted) and $T=0.42$ (gray solid). $R_{12}=0.5$.}
\label{TRB_temp_abs}
\end{figure}

\begin{figure}[!tbh]
  %\centerline{\includegraphics[clip=true,width=3.1in]{fig3.eps}}
            %\centerline{\includegraphics[clip=true,width=2.5in]{fig1b.eps}}
   \begin{minipage}[b]{\linewidth}
     \centerline{\includegraphics[clip=true,width=2.7in]{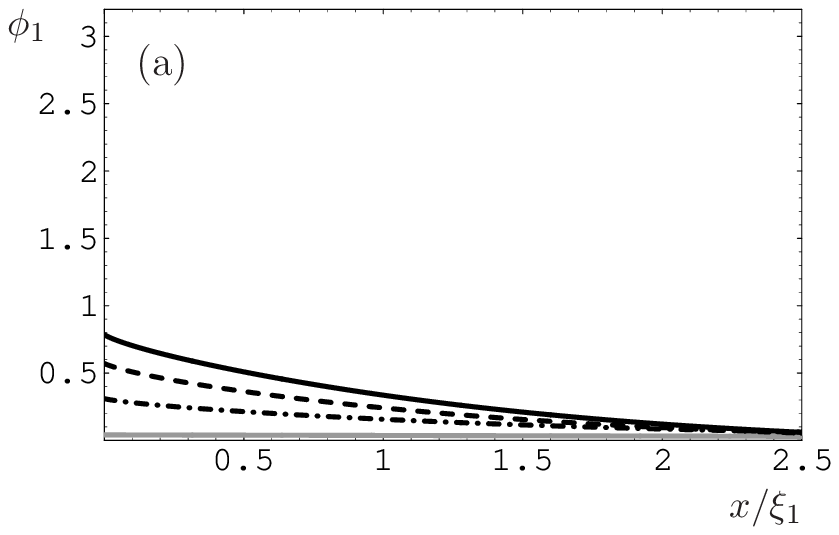}}
     \end{minipage}\hfill
    \begin{minipage}[b]{\linewidth}
   \centerline{\includegraphics[clip=true,width=2.7in]{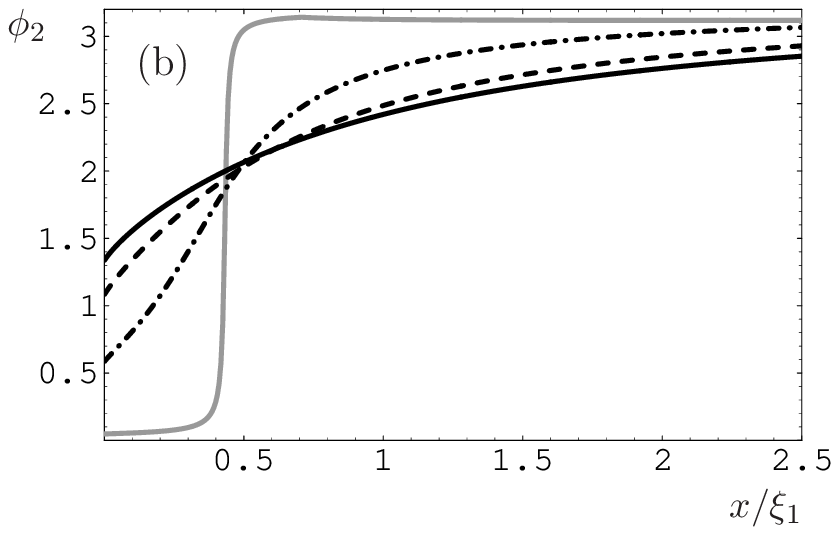}}
  \end{minipage}
   \caption{The OP phases $\phi_1$ (a) and $\phi_2$ (b) as functions of the spatial coordinate $x$. 
The temperatures and the other parameters are the same as in Fig.~\ref{TRB_temp_abs}.}
\label{TRB_temp_phase}
\end{figure}

The temperature evolution of the TRB solution is represented in Figs.~\ref{TRB_temp_abs} and \ref{TRB_temp_phase}. 
Fig.~\ref{TRB_temp_abs} shows the OP absolute values, and Fig.~\ref{TRB_temp_phase} demonstrates the evolution
of the OP phases. It is seen from
the Figures that there is a rather narrow temperature region from $\approx 0.3T_c$ to
$\approx 0.4T_c$, where the OP profiles
change qualitatively from the TRB state with nonzero phase to the real TR solution. The TRB state
completely disappears above $T \sim 0.42T_c$. At the same time the TR solution does not manifest
any qualitative changes in the whole temperature region from $0$ to $T_c$.

\begin{figure}[!tbh]
  \centerline{\includegraphics[clip=true,width=3in]{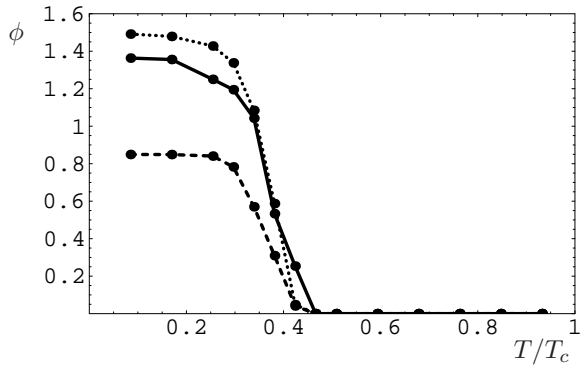}}
            %\centerline{\includegraphics[clip=true,width=2.5in]{fig1b.eps}}
      \caption{OP phases $\phi_{1,2}$ for the TRB state at the surface $x=0$ as functions of temperature.
Dashed line: $\phi_1(x=0)$ for $R_{12}=0.5$, dotted line: $\phi_2(x=0)$ for $R_{12}=0.5$
and black solid line: $\phi_1(x=0)=\phi_2(x=0)$ for $R_{12}=1$.}
\label{phase_temp}
\end{figure}

In order to more explicitly show the temperature transition between the TRB and the TR states
in Fig.~\ref{phase_temp} we plot the OP phases $\phi_{1,2}$ for the TRB state
at the surface $x=0$ as functions of temperature.
The dashed and dotted lines represent phases $\phi_1(x=0)$ and $\phi_2(x=0)$, respectively, calculated
for $R_{12}=0.5$. The black solid line corresponds to the limit of the full interband scattering $R_{12}=1$.
In this case the values of $\phi_1(x=0)$ and $\phi_2(x=0)$ approximately coincide.
The sharp conversion of the TRB
surface state to the TR $s_{++}$ surface state is clearly seen at $T \approx 0.4 T_c$.
It is important that we consider the case of weak interband coupling $|\lambda_{12}| \ll |\lambda_{11}|,|\lambda_{22}|$.
Such a choice of parameters is consistent with the experimental estimates of the coupling constants
for FeSe \cite{khasanov10}. For more strong interband coupling the temperature transition between
the TRB and the TR states would be more smeared.

\begin{figure}[!tbh]
  \centerline{\includegraphics[clip=true,width=2.8in]{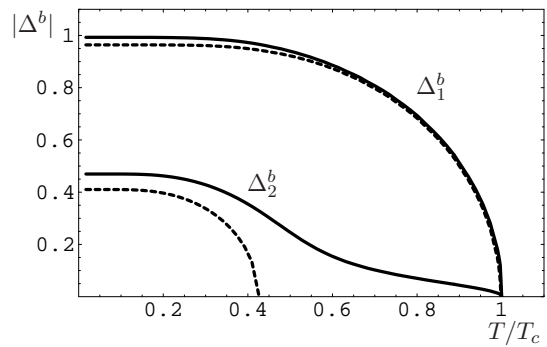}}
            %\centerline{\includegraphics[clip=true,width=2.5in]{fig1b.eps}}
      \caption{Bulk values $\Delta_{1}^b$ and $\Delta_2^b$ as functions of temperature
      for the system under consideration.
The dashed lines show $\Delta_{1}^b$ and $\Delta_2^b$
under the condition of no interband coupling $\lambda_{12}=0$.}
\label{delta_bulk_T}
\end{figure}

For the regime of weak interband coupling there are simple physical arguments indicating
that the TRB state at low temperatures is more energetically favorable than the TR one.
Let us consider the spatial region, where the OP $\Delta_2(x)^{real}$, corresponding to
the TR solution crosses zero point.
It is the region giving the main difference between the free energies of the TRB and the TR states.
Let us imagine that $\Delta_2(x)^{real}$ acquires small constant imaginary part $i \delta \Delta_2$
in the vicinity of the zero crossing point. Then, it is obvious that the leading (quadratic) term of
GL free energy $F \sim a_1 |\Delta_1|^2 + a_2 |\Delta_2|^2 + b (\Delta_1^* \Delta_2+\Delta_1 \Delta_2^*)$
decreases for $T<T_{c2}$ due to the negative contribution of the term $a_2 |\Delta_2|^2$ and increases
for $T>T_{c2}$ due to the positive contribution of this term, while the coupling term remains unaffected
by the small imaginary part. Here $T_{c2}$ is the bulk critical temperature of the lesser order parameter
$\Delta_2$ under the condition of no interband coupling $\lambda_{12}=0$ (See Fig.~\ref{delta_bulk_T}).
Therefore, this rough consideration indicates that the TR state is unstable at low temperatures.
It is the TRB state that becomes energetically favorable at $T<T_{c2}$.
Indeed, from Fig.~\ref{phase_temp} it is seen that the TRB state disappears
at $T \approx T_{c2}$. For the regime of strong interband coupling all the above arguments
are hardly applicable and in order to conclude which state (TRB or TR) is more favorable
the more rigorous consideration of free energy functional is necessary. However, we do not consider
the strong interband coupling regime in the paper.

\section{LDOS}

\label{bs}

\begin{figure*}[!tbh]
  %\centerline{\includegraphics[clip=true,width=3.1in]{fig3.eps}}
            %\centerline{\includegraphics[clip=true,width=2.5in]{fig1b.eps}}
   \begin{minipage}[b]{.33\linewidth}
     \centerline{\includegraphics[clip=true,width=1.8in]{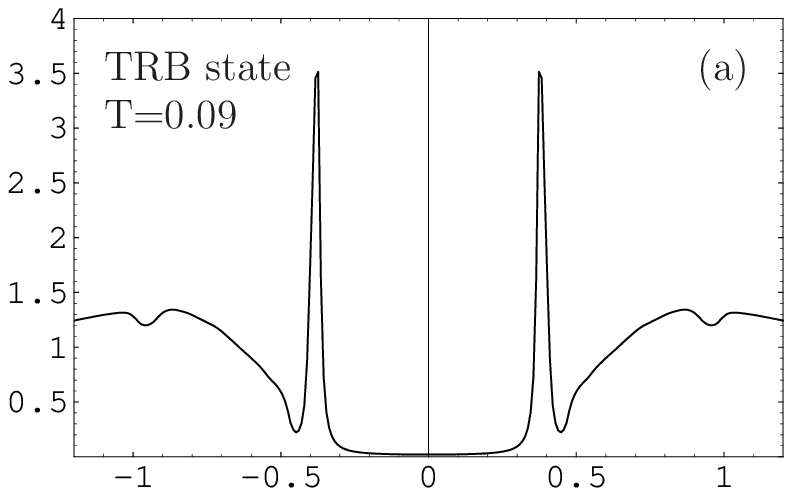}}
     \end{minipage}\hfill
    \begin{minipage}[b]{.33\linewidth}
   \centerline{\includegraphics[clip=true,width=1.8in]{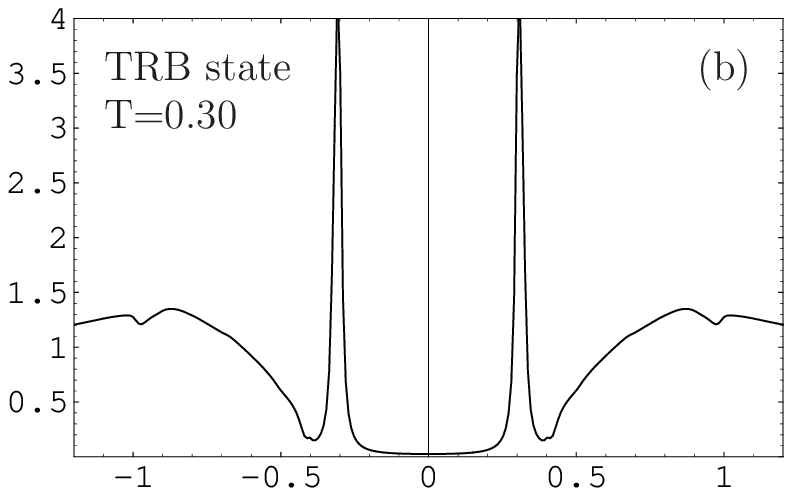}}
  \end{minipage}
\begin{minipage}[b]{.33\linewidth}
     \centerline{\includegraphics[clip=true,width=1.8in]{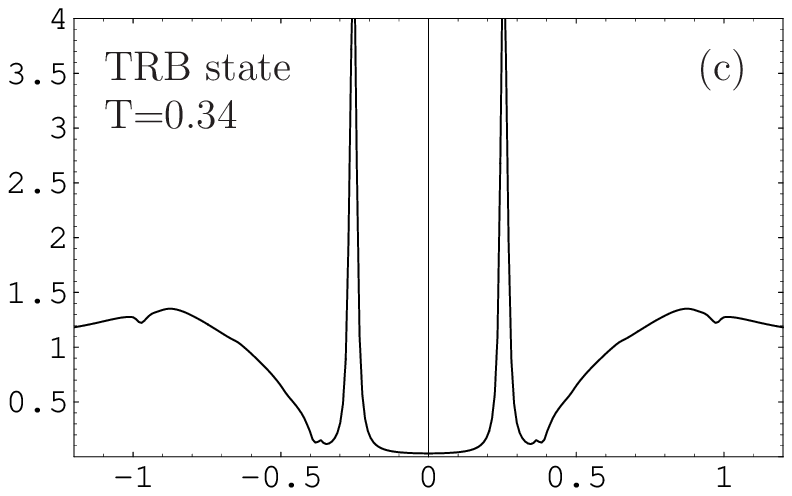}}
     \end{minipage}\hfill
    \begin{minipage}[b]{.33\linewidth}
   \centerline{\includegraphics[clip=true,width=1.8in]{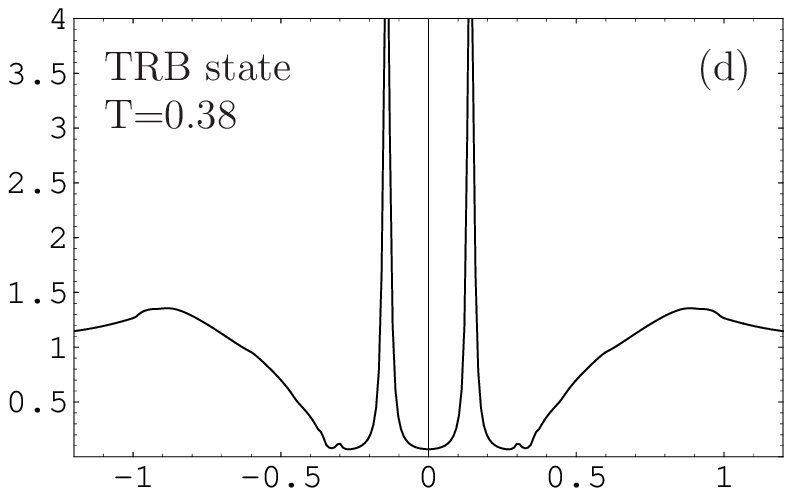}}
  \end{minipage}
\begin{minipage}[b]{.33\linewidth}
     \centerline{\includegraphics[clip=true,width=1.8in]{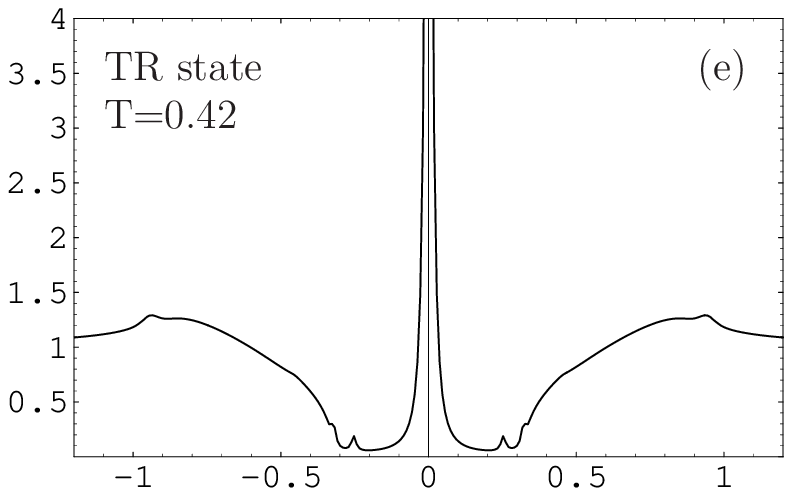}}
     \end{minipage}\hfill
    \begin{minipage}[b]{.33\linewidth}
   \centerline{\includegraphics[clip=true,width=1.8in]{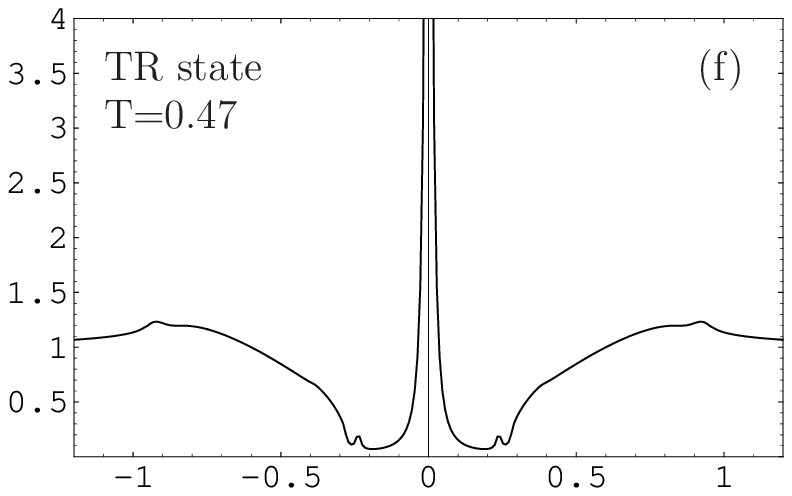}}
  \end{minipage}
   \caption{LDOS at the surface $x=0$ as a function of quasiparticle energy $\varepsilon$, which is measured
in units of $\Delta_1^b(T=0)$. The case of full interband scattering $R_{12}=1$.}
\label{LDOS_R_1}
\end{figure*}

Now we discuss how the described above OP spatial behavior affects the LDOS
near the surface. LDOS $\rho(x,\varepsilon)$ is calculated via the coherence functions as follows
\begin{equation}
\rho(x,\varepsilon)=\sum \limits_i {\rm Re} \left \langle
\frac{1-\gamma_{i\bm p_f}\tilde \gamma_{i \bm p_f}}
{1+\gamma_{i\bm p_f}\tilde \gamma_{i \bm p_f}} \right \rangle_{\bm p_f}
\label{LDOS}
\enspace .
\end{equation}

As it was already discussed in the literature, if $R_{12} \neq 0$ there are surface
bound states in the system, which manifest themselves as well-pronounced peaks in the LDOS.
We begin by discussing the case of pure interband scattering $R_{12} \to 1$. Although this limit can
hardly be realized in real experiments, the influence of the surface OP state on the LDOS is the most striking
here. Then we turn to discussion of the effects taking place at more realistic intermediate values of
the interband scattering.

It was found \cite{bobkov10} that for the TR solution at $R_{12} \to 1$ the bound state
energy tends to zero. For this case the LDOS is dominated by very strong zero-energy peak.
If the TRB surface state is realized in the system, then this is not the case. Instead of well-pronounced zero-energy
peak the LDOS is characterized by two symmetric finite-energy bound state peaks, which are very close
to the smaller gap edges. This remarkable fact can be used in experiment in order to answer the question
if there is the temperature transition between the TRB and the TR $s_{++}$ surface states in the system.
Indeed, if the TRB is realized at low temperature, then at some intermediate temperature it should evolve rather
sharply to the TR $s_{++}$ surface state. This transition is accompanied by the corresponding sharp modification
of the LDOS spectra: two symmetric finite energy peaks divided by U-shaped inner gap evolve into the zero-bias
peak. This is demontrated in Fig.~\ref{LDOS_R_1}. It is worth noting that the main evolution of the LDOS
occurs in a rather narrow temperature interval, where the transition between the TRB and the $s_{++}$ surface states
takes place. The transition region is represented in panels (b)-(e) of Fig.~\ref{LDOS_R_1}. Beyond this temperature
interval [panels (a) and (f)] the modification of the LDOS is not qualitative and mainly connected to the temperature
evolution of the bulk superconducting gaps.

\begin{figure}[!tbh]
  %\centerline{\includegraphics[clip=true,width=3.1in]{fig3.eps}}
            %\centerline{\includegraphics[clip=true,width=2.5in]{fig1b.eps}}
   \begin{minipage}[b]{.5\linewidth}
     \centerline{\includegraphics[clip=true,width=1.65in]{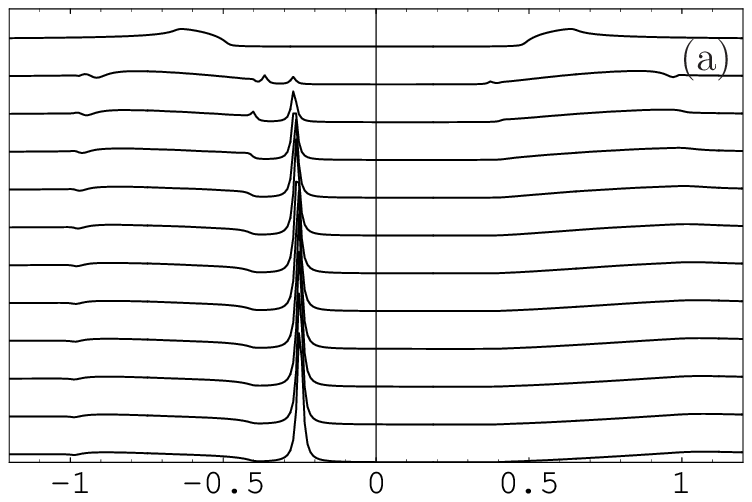}}
     \end{minipage}\hfill
    \begin{minipage}[b]{.5\linewidth}
   \centerline{\includegraphics[clip=true,width=1.65in]{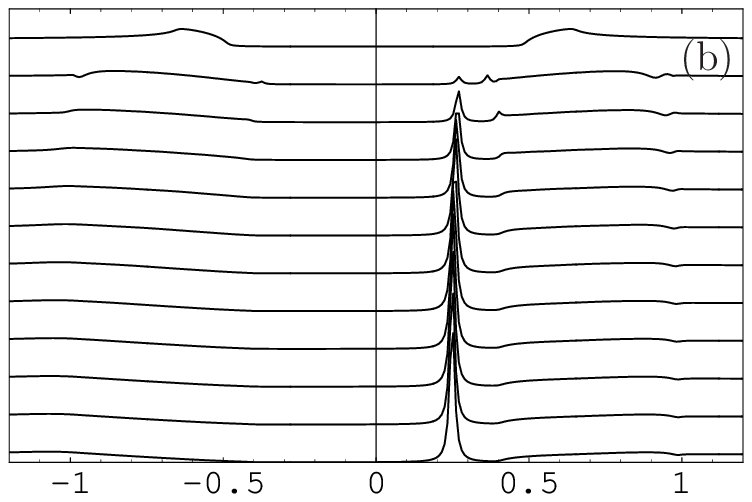}}
  \end{minipage}
\begin{minipage}[b]{.5\linewidth}
     \centerline{\includegraphics[clip=true,width=1.65in]{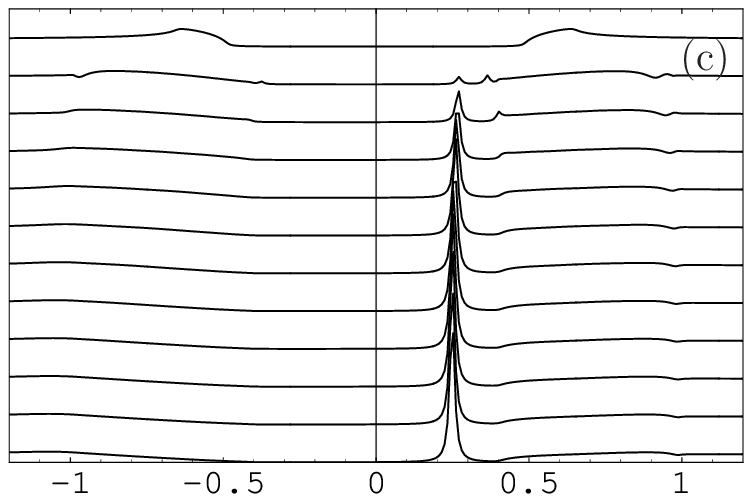}}
     \end{minipage}\hfill
    \begin{minipage}[b]{.5\linewidth}
   \centerline{\includegraphics[clip=true,width=1.65in]{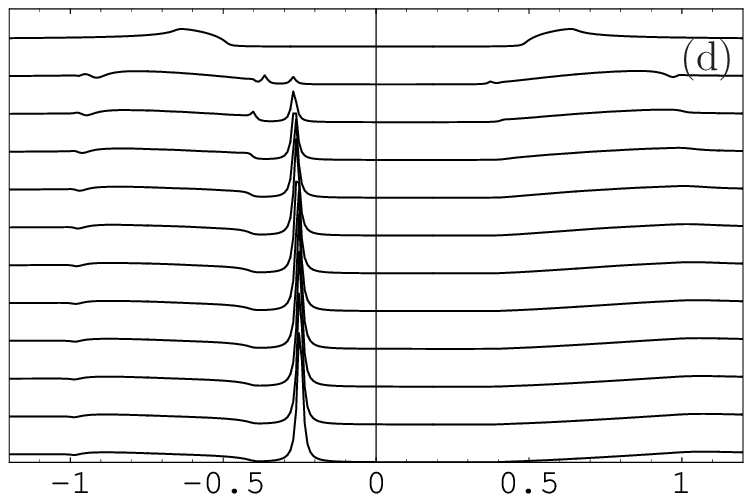}}
  \end{minipage}
   \caption{Angle-resolved density of states at $x=0$ for the TRB state as a function of quasiparticle energy
$\varepsilon$. For all the panels: different curves correspond to
different values of the angle $\theta$ between the quasiparticle trajectory and the normal to the surface.
$\theta$ increases from bottom to top. (a) incoming trajectories for band 1; (b) outgoing trajectories
for band 1; (c) incoming trajectories for band 2; (d) outgoing trajectories for band 2. Full interband scattering
$R_{12}=1$, $T=0.34$.}
\label{spectra_R_1}
\end{figure}

\begin{figure*}[!tbh]
  %\centerline{\includegraphics[clip=true,width=3.1in]{fig3.eps}}
            %\centerline{\includegraphics[clip=true,width=2.5in]{fig1b.eps}}
   \begin{minipage}[b]{.33\linewidth}
     \centerline{\includegraphics[clip=true,width=1.8in]{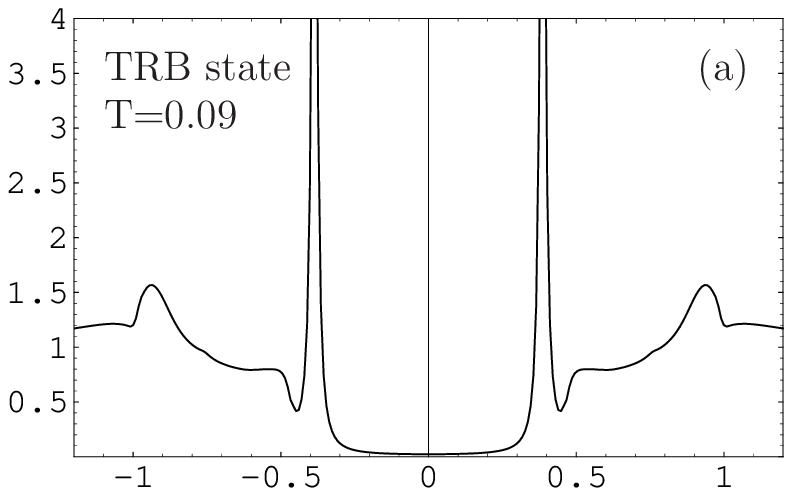}}
     \end{minipage}\hfill
    \begin{minipage}[b]{.33\linewidth}
   \centerline{\includegraphics[clip=true,width=1.8in]{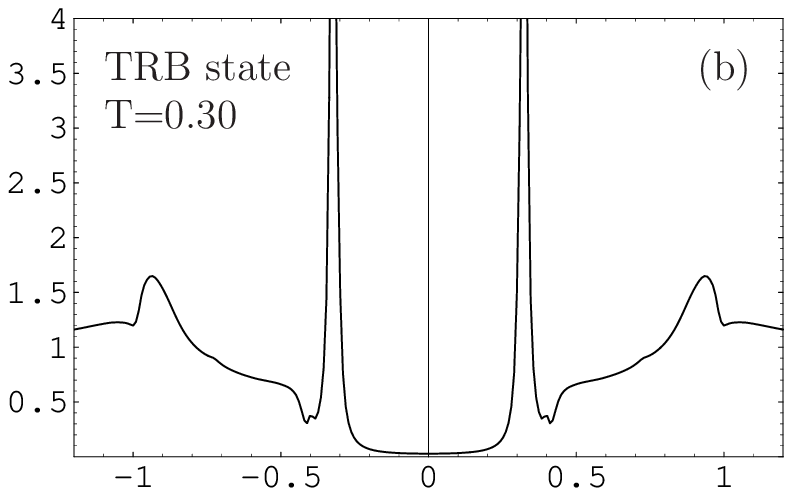}}
  \end{minipage}
\begin{minipage}[b]{.33\linewidth}
     \centerline{\includegraphics[clip=true,width=1.8in]{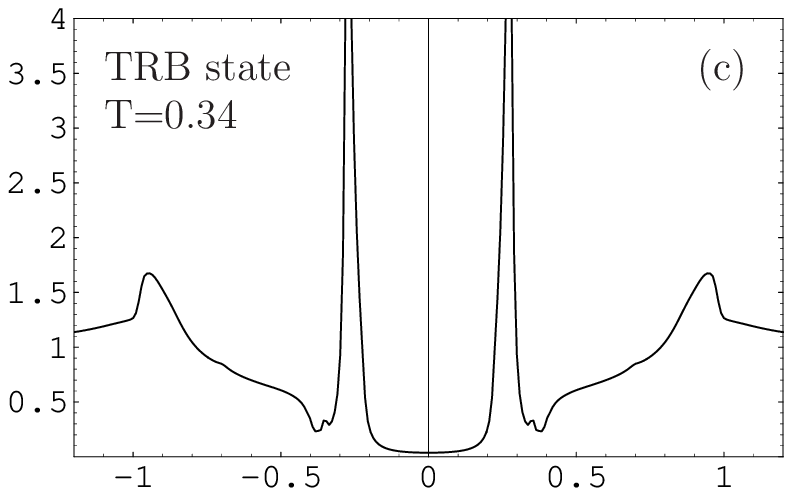}}
     \end{minipage}\hfill
    \begin{minipage}[b]{.33\linewidth}
   \centerline{\includegraphics[clip=true,width=1.8in]{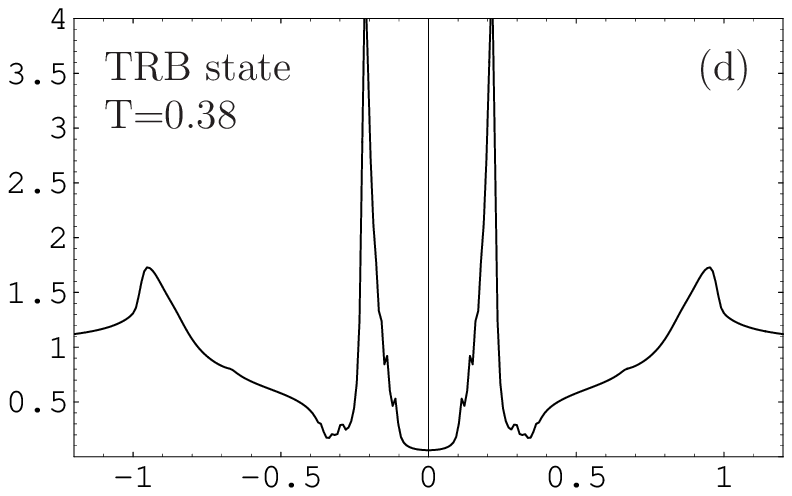}}
  \end{minipage}
\begin{minipage}[b]{.33\linewidth}
     \centerline{\includegraphics[clip=true,width=1.8in]{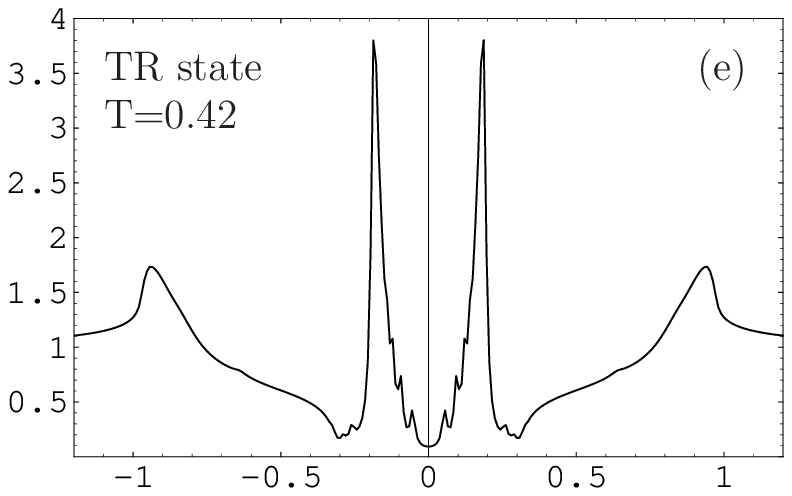}}
     \end{minipage}\hfill
    \begin{minipage}[b]{.33\linewidth}
   \centerline{\includegraphics[clip=true,width=1.8in]{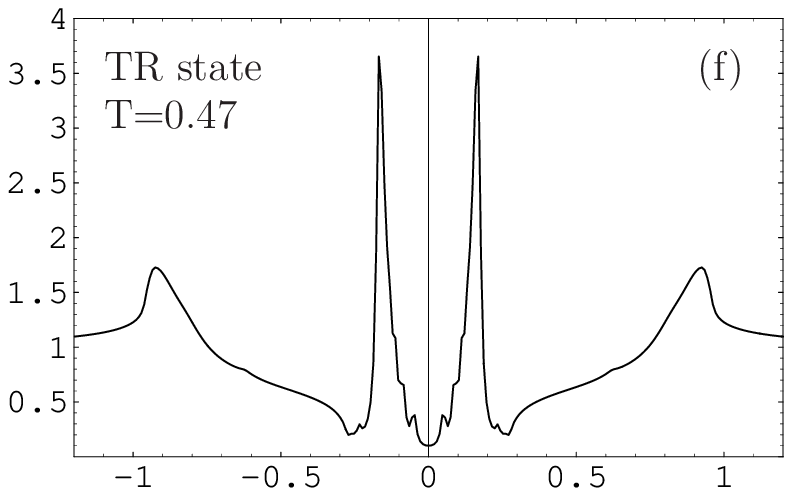}}
  \end{minipage}
   \caption{LDOS at the surface $x=0$ as a function of quasiparticle energy $\varepsilon$.
The case of intermediate interband scattering $R_{12}=0.5$.}
\label{LDOS_R_05}
\end{figure*}

The LDOS behavior typical for the TRB state and demonstrated in panels (a)-(c) of Fig.~\ref{LDOS_R_1}
is quite similar to the LDOS spectra corresponding to the ordinary $s_\pm$ surface state, where the bulk
$s_{\pm}$ OP is only suppressed to some extent near the surface. This surface state should be realized
for weak enough interband scattering $R_{12} \lesssim 0.2$ \cite{bobkov10}. Having in mind that
it is very hard to estimate the value of the interband scattering $R_{12}$ for the particular experimental
situation the question emerges: is it possible to conclude on the basis of the LDOS spectra which surface state
(suppressed $s_\pm$ or TRB) takes place in the system? The answer can be obtained from observation the temperature
evolution of the LDOS: while for the TRB state the discussed above sharp modification of the LDOS should be observed,
there should be no qualitative modification of the LDOS spectra upon increasing of temperature for $s_{\pm}$
surface state.

The second signature of the TRB state can be found in the angle-resolved density of states and is closely connected
to the time-reversal symmetry breaking nature of this state. In Fig.~\ref{spectra_R_1} we demonstrate
the angle-resolved density of states for the case of full interband scattering. Panels (a) and (b) correspond
to incoming and outgoing trajectories, respectively, in band 1, and panels (c) and (d) represent
incoming and outgoing trajectories for band 2. It is clearly seen that the spectra are asymmetric with
respect to zero energy. The reason is the following. The full interband scattering is a special case when
not all the four trajectories (incoming in bands 1 and 2 and outgoing in bands 1 and 2) are involved
in the same scattering event at the surface, but the incoming trajectory from band 1 is only entangled with
the outgoing one from band 2 (the first closed loop) and vice versa: the incoming trajectory from band 2
is only entangled with the outgoing one from band 1 (the second closed loop). These two closed loops
are connected by time-reversal transformation, if one disregards the difference between $\bm p_{||}$
and $-\bm p_{||}$ (there are no reasons for such a difference because we do not study any effects of spin-orbit
interaction and so on). The energies of the bound states, living on these loops, are connected by
sign change. This is a manifestation of the TRB nature of the surface state. It is worth noting here that
the substitution $\phi_{1,2} \to -\phi_{1,2}$, that is the choice of the conjugated OP solution, leads to
the sign change of all the bound state peaks in Fig.~\ref{spectra_R_1}. However, integration over
the Fermi surface results in disappearance of the difference between the two possible OP solutions in the LDOS.

\begin{figure}[!tbh]
  %\centerline{\includegraphics[clip=true,width=3.1in]{fig3.eps}}
            %\centerline{\includegraphics[clip=true,width=2.5in]{fig1b.eps}}
   \begin{minipage}[b]{.5\linewidth}
     \centerline{\includegraphics[clip=true,width=1.65in]{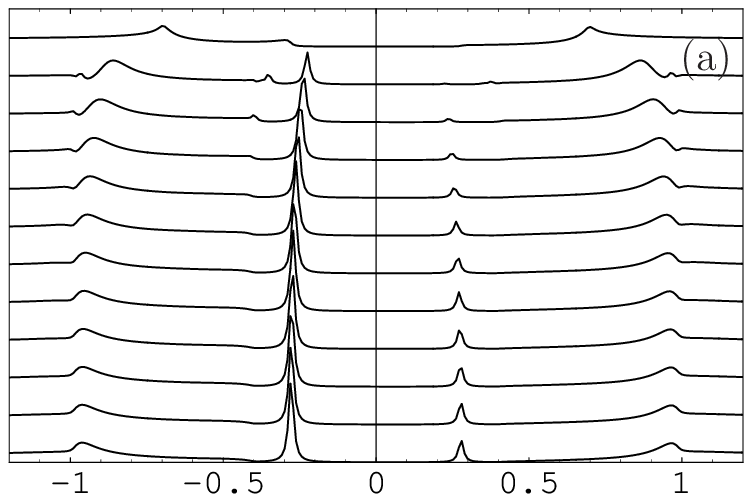}}
     \end{minipage}\hfill
    \begin{minipage}[b]{.5\linewidth}
   \centerline{\includegraphics[clip=true,width=1.65in]{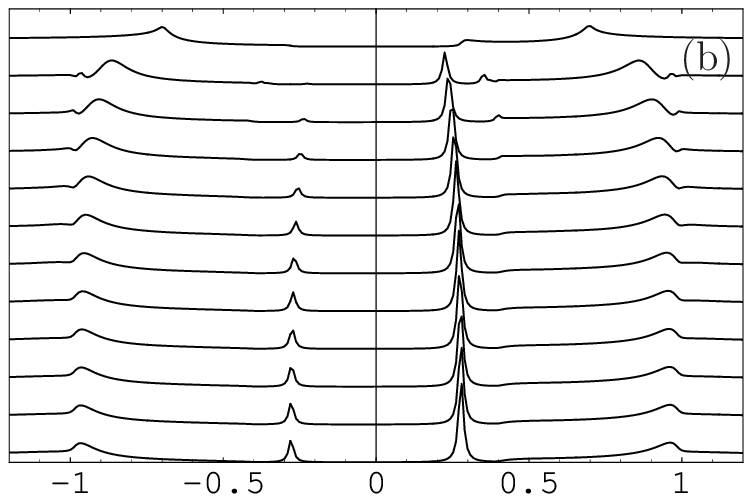}}
  \end{minipage}
\begin{minipage}[b]{.5\linewidth}
     \centerline{\includegraphics[clip=true,width=1.65in]{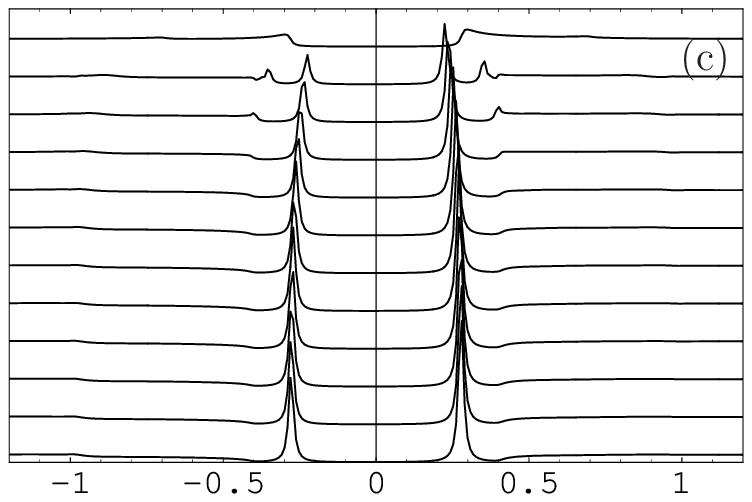}}
     \end{minipage}\hfill
    \begin{minipage}[b]{.5\linewidth}
   \centerline{\includegraphics[clip=true,width=1.65in]{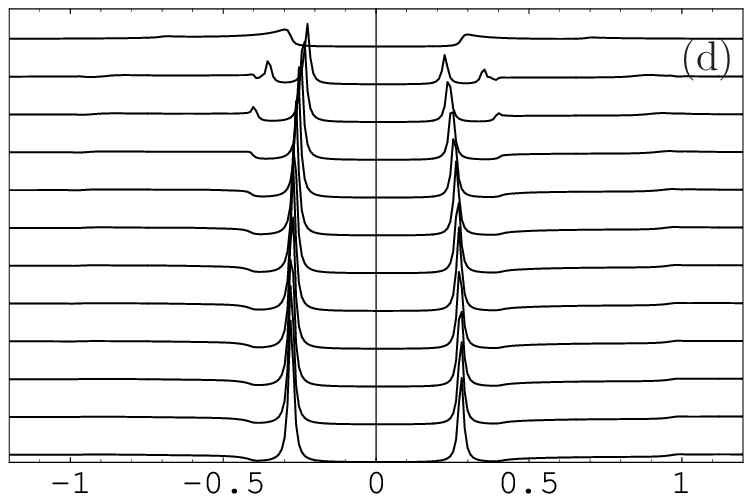}}
  \end{minipage}
   \caption{The same as in Fig.~\ref{spectra_R_1}, but for the intermediate interband scattering $R_{12}=0.5$.}
\label{spectra_R_05}
\end{figure}

Now we turn to the more realistic case of the intermediate interband scattering $R_{12}=0.5$.
For the intermediate value of interband scattering there are finite energy bound states in the system,
which merge with the smaller gap edges when $R_{12} \to 0$. The LDOS behavior for the low-temperature
TRB state and its evolution upon transition to the TR $s_{++}$ surface state is demonstrated
in Fig.~\ref{LDOS_R_05}. Although the zero-energy peak is absent for $R_{12}<1$, the qualitative
difference between the LDOS spectra in the TRB and the $s_{++}$ state exists. While in the TRB state
the bound state peaks are divided by the clearly defined U-shaped gap, the transition to the $s_{++}$ state
results in transforming this U-shaped gap into V-shaped behavior,
which is known to be more typical for the superconductors with OP nodes at the Fermi surface.
As it was shown \cite{bobkov10}, this fact is a consequence of
spatial line of OP nodes appearing if the OP sign reversal takes place near the surface.
Exactly as it was for the case of full interband scattering, the main qualitative evolution of the LDOS
occurs in a rather narrow temperature interval, where the transition between the TRB and the $s_{++}$ surface states
takes place.

In this case angle-resolved spectra, calculated for the TRB state, also manifest the energy asymmetry. However,
in contrast to the limit of full interband scattering, now the both bound states (corresponding to
positive and negative energies) are present for all the trajectories,
but the corresponding peaks in the density of states have asymmetric heights. This is a consequence of the fact
that for the intermediate interband scattering all four trajectories are involved in each scattering event
at the surface.

It is seen from Figs.~\ref{spectra_R_1} and \ref{spectra_R_05} that the bound states are practically dispersionless
for $R_{12}=1$ and are only slightly dispersive for $R_{12}=0.5$. This is a consequence of our simplified model for
the surface: we assume that the scattering matrix elements do not depend on the direction of
of the incoming quasiparticle momentum. In order to check the generality of our main conclusions
we have considered a number of momentum-dependent models for the scattering matrix. We have found that, although
this leads to more dispersive angle-resolved spectra, the main qualitative effects, concerning the sharp
modification of the LDOS upon varying temperature and the asymmetric angle-resolved spectra in the TRB state,
are not sensitive to the details of the particular model.

\section{conclusions}

\label{conc}

On the basis of the self-consistent quasiclassical approach we have theoretically investigated
which types of superconducting order parameters are possible near the surface of a two-band superconductor
with $s_{\pm}$ order parameter in the bulk. It is found that the answer is mainly determined by
the value of interband surface scattering with respest to the intraband one. If the intraband
surface scattering $R_0$ dominates the interband one $R_{12}$, that is $R_0$ is close to unity,
then the superconducting OP is only suppressed to some extent near the surface for the both bands.
For intermediate and strong interband scattering there are two possible non-trivial surface
states of the order parameter. Further the particular answer depends on the ratio between the intraband
and interband pairing constants. We have focused on the case of weak interband coupling.
It is found that at low temperatures the self-consistency equation has two solutions: (i)
purely real solution, where the symmetry of the superconducting state near the
surface is changed from $s_{\pm}$ to conventional $s_{++}$. In this state the OP belonging to
band 2 (with the smaller absolute value of the bulk OP) changes its sign near the surface via
crossing zero value at some distance from the surface. (ii) Time-reversal symmetry breaking solution:
upon approaching the surface the order parameters in the two bands acquire
phases $\phi_{1,2}(x) \neq (0,\pi)$, while the bulk values $\phi_1 = 0$ and $\phi_2 = \pi$ restore
within a few superconducting coherence lengths from the surface.
We argue that at low temperatures the TRB surface state can be more energetically favorable
than the $s_{\pm} \to s_{++}$ time reversal symmetry conserving state.
For higher temperatures up to $T_c$ only the TR state can exist. The transition between
the two temperature regions is rather sharp.

Further, we show that the local density of states is very sensitive to the particular surface
state of the order parameter and, therefore, experience a sharp change upon varying temperature.
For the case of full interband scattering the finite-energy peaks, which are
located close to the smaller gap edges and are typical for the TRB state, are changed by the
zero-energy peak upon increasing of the temperature. For the case of intermediate interband scattering
the zero-energy peak cannot exist, but the U-shaped low-energy LDOS behavior is changed by the V-shaped
LDOS behavior at the transition from the TRB to the $s_{++}$ surface state. In addition, the angle-resolved
spectra manifest the energy asymmetry in the TRB state. Therefore, the signatures of the transition between the TRB
and the $s_{++}$ surface states can be detected by the density of states measurements.

%\begin{figure}[!tbh]
  %\centerline{\includegraphics[clip=true,width=3.1in]{fig3.eps}}
            %\centerline{\includegraphics[clip=true,width=2.5in]{fig1b.eps}}
   %\begin{minipage}[b]{.5\linewidth}
    % \centerline{\includegraphics[clip=true,width=1.651in]{fig3a_0.eps}}
     %\end{minipage}\hfill
    %\begin{minipage}[b]{.5\linewidth}
   %\centerline{\includegraphics[clip=true,width=1.7in]{fig3b_0.eps}}
  %\end{minipage}
   %\caption{}
%\label{Delta_lambda_intra}
%\end{figure}

\begin{acknowledgments}
The authors would like to thank A.S. Mel'nikov and M.A. Silaev for fruitful discussions.
The support by RFBR Grant 09-02-00799 is acknowledged.
\end{acknowledgments}

%\bibliography{/users/tkm/howell/latexx/bibtexx/refs}

\end{document}